\documentclass[prl,twocolumn,floatix,amssymb,showpacs,amsmath,superscriptaddress]{revtex4-1}
\usepackage{graphicx}
\usepackage{epsfig}
\usepackage{amsmath,amssymb}

\usepackage{color}
\usepackage{tikz}
\usepackage[T1]{fontenc}

\begin{document}

\newcommand{\ket}[1]{\ensuremath{\left|{#1}\right\rangle}}
\newcommand{\bra}[1]{\ensuremath{\left\langle{#1}\right|}}
\newcommand{\braket}[2]{\ensuremath{\langle{#1}|{#2}\rangle}}
\newcommand{\brakket}[3]{\ensuremath{\langle{#1}|{#2}|{#3}\rangle}}
\newcommand{\quadr}[1]{\ensuremath{{\not}{#1}}}
\newcommand{\quadrd}[0]{\ensuremath{{\not}{\partial}}}
\newcommand{\slpar}{\partial\!\!\!/}
\newcommand{\gtrescero}{\gamma_{(3)}^0}
\newcommand{\gtresuno}{\gamma_{(3)}^1}
\newcommand{\gtresi}{\gamma_{(3)}^i}

\title{Digital Quantum Rabi and Dicke Models in Superconducting Circuits}

\date{\today}

\author{A. Mezzacapo}
\affiliation{Department of Physical Chemistry, University of the Basque Country
UPV/EHU, Apartado 644, E-48080 Bilbao, Spain}
\author{U. Las Heras}
\affiliation{Department of Physical Chemistry, University of the Basque Country
UPV/EHU, Apartado 644, E-48080 Bilbao, Spain}
\author{J. S. Pedernales}
\affiliation{Department of Physical Chemistry, University of the Basque Country
UPV/EHU, Apartado 644, E-48080 Bilbao, Spain}
\author{L. DiCarlo}
\affiliation{Kavli Institute of Nanoscience, Delft University of Technology, P. O. Box 5046, 2600 GA Delft, The Netherlands}
\author{E. Solano}
\affiliation{Department of Physical Chemistry, University of the Basque Country
UPV/EHU, Apartado 644, E-48080 Bilbao, Spain}
\affiliation{IKERBASQUE, Basque Foundation for Science, Alameda Urquijo 36, 48011
Bilbao, Spain}
\author{L. Lamata} 
\affiliation{Department of Physical Chemistry, University of the Basque Country
UPV/EHU, Apartado 644, E-48080 Bilbao, Spain}

\begin{abstract}
We propose the analog-digital quantum simulation of the quantum Rabi and Dicke models using circuit quantum electrodynamics (QED). We find that all physical regimes, in particular those which are impossible to realize in typical cavity QED setups, can be simulated via unitary decomposition into digital steps. Furthermore, we show the emergence of the Dirac equation dynamics from the quantum Rabi model when the mode frequency vanishes. Finally, we analyze the feasibility of this proposal under realistic superconducting circuit scenarios.\end{abstract}

\pacs{03.67.Lx, 42.50.Pq, 02.30.Ik}

\maketitle

The simplest, most fundamental interaction of quantum light and quantum matter can be described by the quantum Rabi model, consisting of the dipolar coupling of a two-level system with a single radiation mode~\cite{Rabi36}. The Dicke model~\cite{Dicke54} was later introduced to generalize this interaction to an ensemble of $N$ two-level systems. Typically, the coupling strength is small compared to the transition frequencies of the two-level system and the radiation mode, leading to effective Jaynes-Cummings and Tavis-Cummings interactions, respectively, after performing a rotating-wave approximation (RWA). This introduces a $U(1)$ symmetry and integrability to the model for any $N$~\cite{Jaynes63,Tavis68}. Recently, analytical solutions for the generic quantum Rabi and Dicke models for $N=3$ were found~\cite{Braak11,BraakDicke3}. However, the general case for arbitrary $N$ is still unsolved, while its direct study in a physical system remains an outstanding challenge. 

A variety of quantum platforms, such as cavity QED, trapped ions, and circuit QED, provides a natural implementation of the Jaynes-Cummings and Tavis-Cummings models, due to the weak qubit-mode coupling strength. When the latter is a fraction or comparable to the mode frequency, the model is said to be in the ultrastrong coupling (USC) regime.  Experimental evidence of this regime has been observed in the optical~\cite{Gunter09} and microwave domains~\cite{Niemczyk10,FornDiaz10}. Coupling strengths larger than the mode frequency mark the transition towards the recently introduced deep-strong coupling (DSC) regime~\cite{Casanova10}. Signatures of the latter may be retrieved effectively in different quantum systems~\cite{Crespi12, Ballester12}, but an experimental observation of the full quantum Rabi and Dicke models in all parameter regimes has not yet been realized. In particular, the quantum simulation~\cite{Feynman82} of the Dicke Hamiltonian could outperform analytical and numerical methods, while enabling the simulation of engineered super-radiant phase transitions~\cite{Hepp73,Wang73,Carmichael73}. Recently, technological improvements of controlled quantum platforms have increased the interest in quantum simulations~\cite{Bloch12,Blatt12,Alan12,Georgescu}. A digital approach to quantum simulations was put forward in Ref.~\cite{Lloyd96}. In this sense, it has been analyzed how suitable versions of digital quantum simulators can be implemented with available quantum platforms~\cite{Lanyon11,Casanova12,Mezzacapo12,Heras}. Standard digital quantum simulations focus on the efficient decomposition of the quantum system dynamics in terms of elementary gates. In order to maximize the efficiency of the simulation, one may analyze which is the decomposition of the dynamics in its largest realizable parts, and reduce the number of elementary interactions in the simulation. This approach can be labeled as analog-digital quantum simulation and corresponds to finding some terms in the simulated system that can be implemented in an analog way, e.g., to employ a harmonic oscillator to simulate a bosonic field, while others will be carried out with a digital decomposition.

In this article, we propose the analog-digital quantum simulation of the quantum Rabi and Dicke models in a circuit QED setup, provided only with Jaynes-Cummings and Tavis-Cummings interactions, respectively. We show how the rotating and counter-rotating contributions to the corresponding dynamics can be effectively realized with digital techniques. By interleaved implementation of rotating and counter-rotating steps, the dynamics of the quantum Rabi and Dicke models can be simulated for all parameter regimes with negligible error. Lastly, we show how a relativistic Dirac dynamics can be retrieved in the limit where the mode frequency cancels.

\section*{Results}
We start by considering a generic circuit QED setup consisting of a charge-like qubit, e.g. a transmon qubit~\cite{Koch07}, coupled to a microwave resonator. The setup is well described by the Hamiltonian ($\hbar=1$)~\cite{Blais04} 
\begin{equation}
H=\omega_r a^{\dagger}a +\frac{\omega_q}{2}\sigma^z +g(a^{\dagger}\sigma^-+a\sigma^+),\label{QubitResHam}
\end{equation}
where $\omega_r$ and $\omega_q$ are the resonator and qubit transition frequencies, $g$ is the resonator-qubit coupling strength, $a^{\dagger}$$(a)$ is the creation(annihilation) operator for the resonator mode, and $\sigma^{\pm}$ raise and lower excitations on the qubit. 
The capacitive interaction in Eq.~(\ref{QubitResHam}) excludes excitations of the higher levels of the qubit device, because typically the coupling $g$ is much smaller than other transition frequencies of the system. By trying to design setups with larger capacitive couplings, pushing them above dispersive regimes, one starts to populate the higher levels of the transmons, producing unwanted leakage. On the other hand, methods based on orthogonal drivings of the qubits~\cite{Ballester12,Pedernales13} may increase the resonator population. Here, we show that the dynamics of the quantum Rabi Hamiltonian 
\begin{equation}
H_R=\omega^R_r a^{\dagger}a +\frac{\omega^R_q}{2}\sigma^z +g^R\sigma^x(a^{\dagger}+a)\label{RabiHam}
\end{equation}
can be encoded in a  superconducting setup provided with a Jaynes-Cummings interaction, as in Eq.~(\ref{QubitResHam}), using a digital expansion. 
\begin{figure}
\includegraphics[scale=0.26]{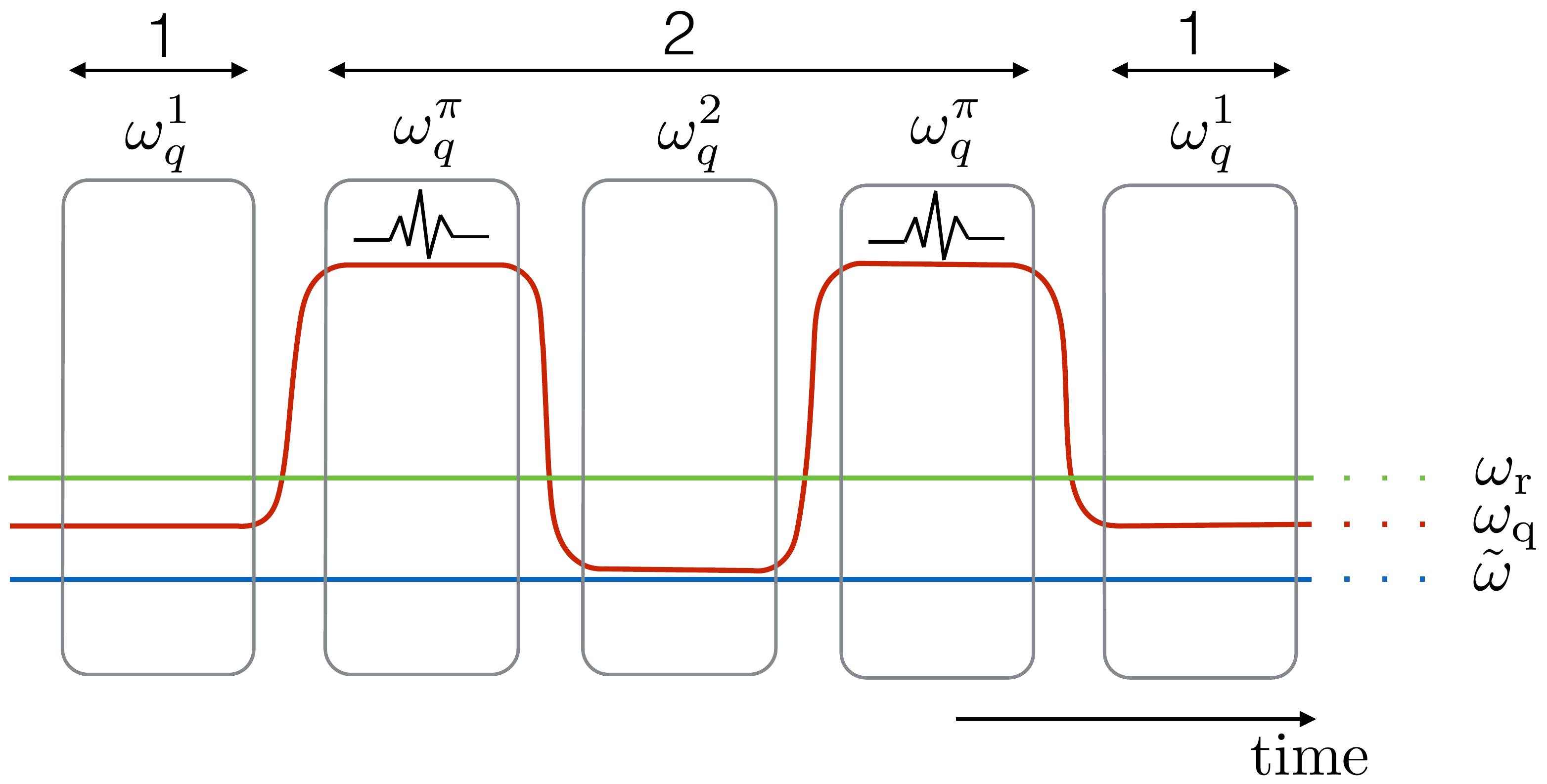}
\caption{(Color online) Frequency scheme of the stepwise implementation for the quantum Rabi Hamiltonian. A transmon qubit of frequency $\omega_q$ is interacting with a microwave resonator, whose transition frequency is $\omega_r$. The interactions $H_{1,2}$ in Eq.~(\ref{Ham12}) are simulated respectively with a Jaynes-Cummings interaction (step 1), and another one with different detuning, anticipated and followed by $\pi$ pulses (step 2). \label{FrequencyScheme}} 
\end{figure}
The quantum Rabi Hamiltonian in Eq.~(\ref{RabiHam}) can be decomposed into two parts, $H_R=H_1+H_2$, where
\begin{eqnarray}
&&H_1=\frac{\omega^R_r}{2} a^{\dagger}a +\frac{\omega^1_q}{2}\sigma^z +g(a^{\dagger}\sigma^-+a\sigma^+) , \nonumber \\
&&H_2=\frac{\omega^R_r}{2} a^{\dagger}a -\frac{\omega^2_q}{2}\sigma^z +g(a^{\dagger}\sigma^++a\sigma^-) ,
\label{Ham12} 
\end{eqnarray}
and we have defined the qubit transition frequency in the two steps such that $\omega_q^1-\omega_q^2=\omega^R_q$. These two interactions can be simulated in a typical circuit QED device with  fast control of the qubit transition frequency. Starting from the qubit-resonator Hamiltonian in Eq.~(\ref{QubitResHam}), one can define a frame rotating at frequency $\tilde{\omega}$, in which the effective interaction Hamiltonian becomes 
\begin{equation}
\tilde{H}=\tilde{\Delta}_ra^{\dagger}a+\tilde{\Delta}_q\sigma^z+g(a^{\dagger}\sigma^-+a\sigma^+),\label{IntHam}
\end{equation}  
with $\tilde{\Delta}_r=(\omega_r-\tilde{\omega})$ and $\tilde{\Delta}_q=\left(\omega_q-\tilde{\omega}\right)/2$. Therefore, Eq.~(\ref{IntHam}) is equivalent to $H_1$, following a proper redefinition of the coefficients.
The counter-rotating term $H_2$ can be simulated by applying a local qubit rotation to $\tilde{H}$ and a different detuning for the qubit transition frequency,
\begin{equation}
e^{-i \pi\sigma^x/2}\tilde{H}e^{i \pi\sigma^x/2}=\tilde{\Delta}_ra^{\dagger}a-\tilde{\Delta}_q\sigma^z+g(a^{\dagger}\sigma^++a\sigma^-).\label{RotHam}
\end{equation}
By choosing different qubit-resonator detuning for the two steps, $\tilde{\Delta}^1_q$ for the first one and $\tilde{\Delta}^2_q$ for the rotated step, one is able to simulate the quantum Rabi Hamiltonian, Eq.~(\ref{RabiHam}), via digital decomposition~\cite{Lloyd96}, by interleaving the simulated interactions. The frequency scheme of the protocol is shown in Fig.~\ref{FrequencyScheme}. Standard resonant Jaynes-Cummings interaction parts with different qubit transition frequencies are interrupted by microwave pulses, in order to perform customary qubit flips~\cite{Blais07}. This sequence can be repeated according to the digital simulation scheme to obtain a better approximation of the quantum Rabi dynamics.

The simulated Rabi parameters can be obtained as a function of the physical parameters of the setup by inverting the derivation presented above. In this way, one has that the simulated bosonic frequency is related to the resonator detuning $\omega_r^R=2\tilde{\Delta}_r$, the two-level transition frequency is related to the transmon frequency in the two steps, $\omega_q^R=\tilde{\Delta}_q^1-\tilde{\Delta}_q^2$, and the coupling to the resonator remains the same, $g^R=g$. Notice that even if the simulated two-level frequency $\omega_q^R$ depends only on the frequency difference, large detunings $\tilde{\Delta}_q^{1(2)}$ will affect the total  fidelity of the simulation. In fact, since the digital error depends on the magnitude of individual commutators between the different interaction steps, using larger detunings linearly increases the latter, which results in fidelity loss of the simulation. To minimize this loss, one can choose, for example, the transmon frequency in the second step to be tuned to the rotating frame, such that $\tilde{\Delta}_q^2=0$. Nevertheless, to avoid sweeping the qubit frequency across the resonator frequency, one may choose larger detunings.
To estimate the loss of fidelity due to the digital approximation of the simulated dynamics, we consider a protocol performed with typical transmon qubit parameters~\cite{Koch07}.
We estimate a resonator frequency of $\omega_r/2\pi=7.5$~GHz, and a transmon-resonator coupling of $g/2\pi=100$~MHz. The qubit frequency $\omega_q$ and the frequency of the rotating frame $\tilde{\omega}$ are varied to reach different parameter regimes.

\begin{table}

\caption{Simulated quantum Rabi dynamics parameters versus frequencies of the system. For all entries in the right column, the resonator frequency is fixed to $\omega_r/2\pi=7.5$~GHz, and the coupling $g^R/2\pi=100$~MHz. Frequencies are shown up to a $2\pi$ factor.\label{Table}}
\vspace{0.2cm}
\begin{tabular}{l|l}
\hline\hline

  $g^R=\omega_q^R/2=\omega_r^R/2$ \;\; & $\tilde{\omega}=7.4$~GHz, $\omega_q^1-\omega_q^2=200$~MHz \;\; \\
  $g^R=\omega_q^R=\omega_r^R$ \;\; & $\tilde{\omega}=7.45$~GHz, $\omega_q^1-\omega_q^2=100$~MHz \;\; \\
  $g^R=2\omega_q^R=\omega_r^R$ \;\; & $\tilde{\omega}=7.475$~GHz, $\omega_q^1-\omega_q^2=100$~MHz \;\; \\

\hline\hline
\end{tabular}
\end{table}
To perform the simulation for the quantum Rabi model with $g^R/2\pi=\omega^R_q/2\pi=\omega^R_r/2\pi=100$~MHz, for example, one can set $\omega^1_q/2\pi=7.55$~GHz, $\omega^2_q/2\pi=7.45$~GHz. In this way, one can define an interaction picture rotating at $\tilde{\omega}/2\pi=7.45$~GHz  to encode the dynamics of the quantum Rabi model with minimal fidelity loss.
Considering that single-qubit rotations take approximately~$\sim10$~ns, tens of Trotter steps could be comfortably performed within the coherence time. Notice that, in performing the protocol, one has to avoid populating the third level of the transmon qubit. Taking into account transmon anharmonicities of about $\alpha=-0.1$, for example, in this case one has third level transition frequencies of $6.795$~GHz and $6.705$~GHz. Therefore, given the large detuning with the resonator, it will not be populated.
Similarly, by choosing different qubit detunings and rotating frames, one can simulate a variety of parameter regimes, e.g. see Table~\ref{Table}.

\section*{Discussion}

\begin{figure}
\includegraphics[scale=0.37]{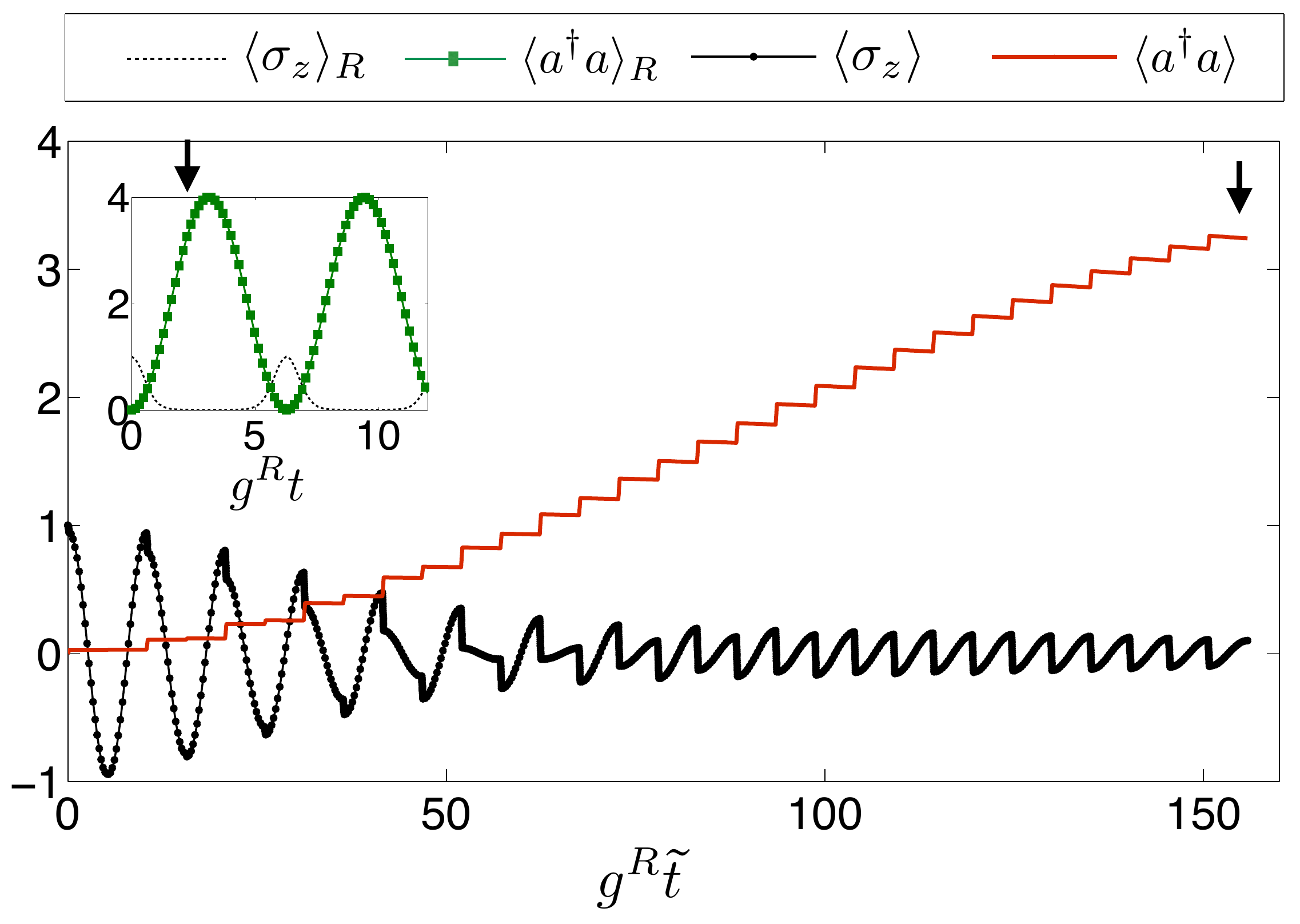}
\caption{ (Color online) A transmon qubit and microwave resonator simulating the quantum Rabi Hamiltonian in the regime $g^R=\omega_r^R$, $\omega_q^R=0$. The ideal dynamics, plotted in the inset, shows collapses and revivals of the photon and qubit population. The latter are recovered via sequential qubit-resonator interactions and qubit flips. The photon population is pumped to the expected value at the time marked by the arrow. Note that the simulating time $\tilde{t}$ is different from the simulated time $t$.  \label{Dynamics}} 
\end{figure}

In order to capture the physical realization of the simulation, we plot in Fig.~\ref{Dynamics} the behavior of the transmon-resonator system during the simulation protocol. We numerically integrate a master equation, alternating steps of Jaynes-Cummings interaction with single-qubit flip pulses. We consider $\dot{\rho}=-i[H,\rho]+\kappa L(a)\rho+\Gamma_\phi L(\sigma^z)\rho+\Gamma_- L(\sigma^-)\rho$, with  Jaynes-Cummings terms $\tilde{H}=\tilde{\Delta}_ra^{\dagger}a+\tilde{\Delta}_q\sigma^z+g(a^{\dagger}\sigma^-+a\sigma^+)$, alternated with qubit-flip operations $H_f=f(t)\sigma^x$, where $f(t)$ is a smooth function such that $\int_0^{T_f}f(t)dt=\pi/2$, $T_f$ being the qubit bit-flip time. The quantum dynamics is affected by Lindblad superoperators $\Gamma_\phi L(\sigma^z)\rho$, $\Gamma_- L(\sigma^-)\rho$, and $\kappa L(a) \rho$ modelling qubit dephasing, qubit relaxation and resonator losses. We have defined $L(A)\rho=(2A\rho A^{\dagger}-A^{\dagger}A\rho-\rho A^{\dagger}A)/2$. We set a resonator-qubit coupling of $g/2\pi=80$~MHz, and a frame rotating at the qubit frequency, $\tilde{\Delta}_q=0$, $\tilde{\Delta}_r/2\pi=40$~MHz. We consider $\Gamma_-/2\pi=30$~kHz, $\Gamma_\phi/2\pi=60$~kHz, and $\kappa/2\pi=100$~kHz. The inset of Fig.~\ref{Dynamics} shows collapses and revivals of both the photon and spin dynamics, which are typical signatures of the regimes of the quantum Rabi dynamics dominated by the coupling strength.  We consider prototypical DSC dynamics, with $\omega_q^R=0$, and $g^R=\omega_r^R$. Notice that to encode the dynamics corresponding to a certain simulated time $t$, one needs the quantum simulator to run for a simulating time $\tilde{t}$, that depends on the specific gate times of the experiment. We choose to set the simulation at the time marked by the black arrow, close to the photon population peak in the inset. A simulation with $15$ digital steps is then performed. The time for a single qubit flip pulse is set to $T_f=10$~ns. Periodic collapses and revivals of the bosonic population of the quantum Rabi model $\langle{a^{\dagger}a}\rangle_R$ are shown as a function of time, in the inset. The ideal spin and bosonic populations $\langle\sigma_z\rangle_R$ and $\langle a^{\dagger}a\rangle_R$, evolving according to the quantum Rabi Hamiltonian, are shown to be in good agreement with the simulated ones, $\langle\sigma_z\rangle$ and $\langle a^{\dagger}a\rangle$, at the final simulated time. In fact, during the Jaynes-Cummings interaction parts, photons are pumped into the resonator. Afterwards, before the photon population starts to decrease due to excitation exchanges with the transmon qubit, a qubit flip further enhances the photon production.

\begin{figure}
\includegraphics[scale=0.37]{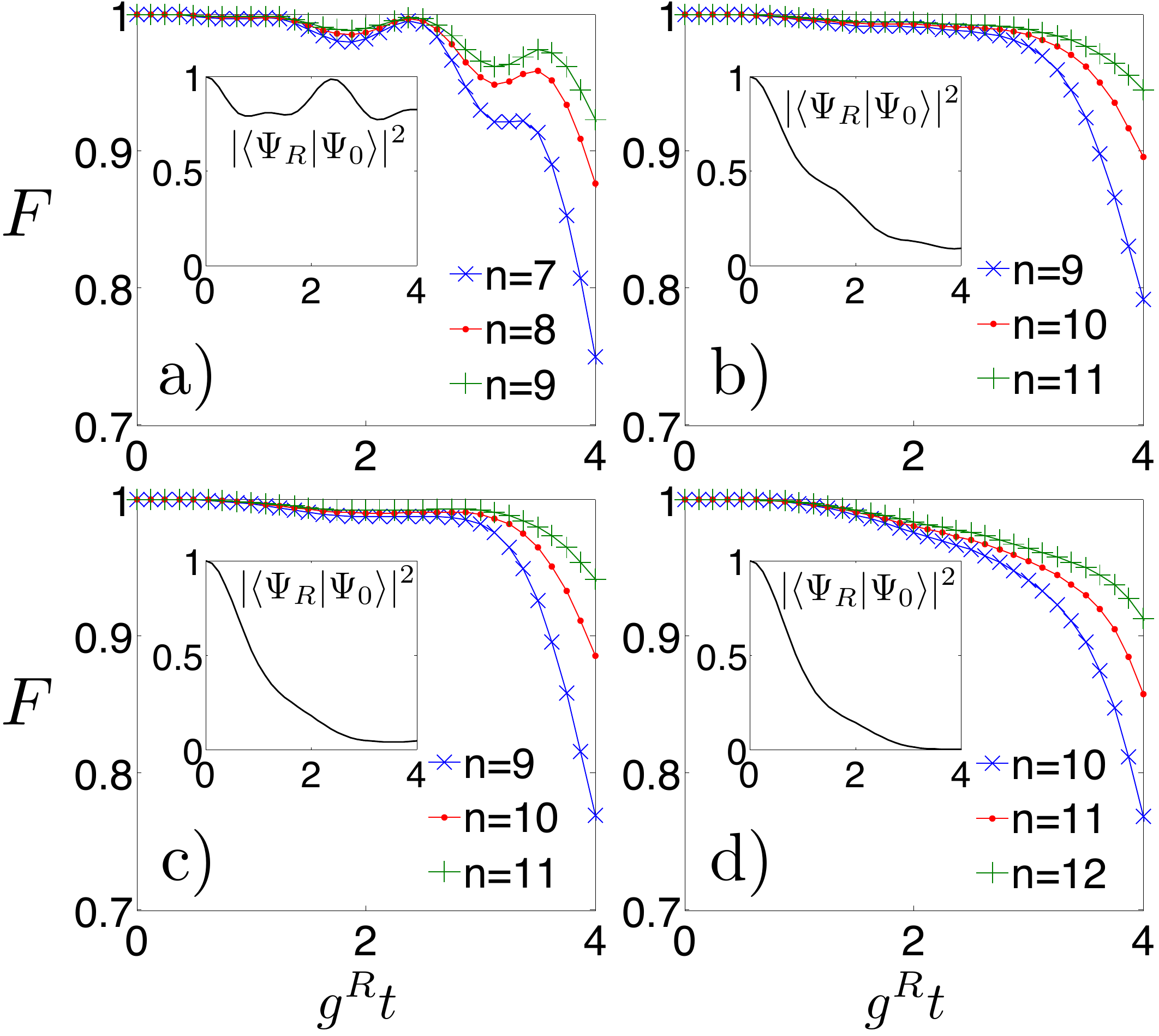}
\caption{ (Color online) Time evolution of the fidelity $F=|\langle\Psi_S|\Psi_R\rangle|^2$ of state $\ket{\Psi_S}$ evolving according to the digitized protocol, to the ideal state $\ket{\Psi_R}$ evolving according to the quantum Rabi dynamics, with a)~$g^R=\omega^R_r/2=\omega^R_q/2$, b)~$g^R=\omega^R_r=\omega^R_q$, c)~$g^R=2\omega^R_r=\omega^R_q$, and d)~$g^R=2\omega^R_r=1.5\omega^R_q$. The simulation is performed for different number $n$ of Trotter steps. Black curves in the insets show the overlap of the ideal evolved state with the one at time $t=0$, $|\langle\Psi_R|\Psi_0\rangle|^2$, initialized with a fully excited qubit and the resonator in the vacuum state. \label{TrotterFidelity}} 
\end{figure}

The simulation protocol can be performed for every time of the dynamics, with the number of digital steps tuned to reach a satisfactory simulation fidelity. We plot in Fig.~\ref{TrotterFidelity} the fidelity $F=|\braket{\Psi_S}{\Psi_R}|^2$ as a function of time of the simulated wavefunction $\Psi_S$, including resonator and spin degrees of freedom, versus the ideal one $\Psi_R$, evolving according to $H_R$, as defined in Eq.~(\ref{RabiHam}). The fidelity is plotted for different parameters and iteration steps. Increasing the number of steps, the fidelity grows as expected from standard Suzuki-Lie-Trotter expansions~\cite{Suzuki90}. In principle, the whole protocol can accurately access non-analytical regimes of these models, including USC and DSC regimes.  

By adding several transmon qubits to the architecture, the presented method can be extended to simulate the Dicke Hamiltonian 
\begin{equation}
H_D=\omega^R_r a^{\dagger}a +\sum_{j=1}^N\frac{\omega^R_q}{2}\sigma_j^z +\sum_{j=1}^Ng^R\sigma^x_j(a^{\dagger}+a).
\end{equation}
This simulation can be efficiently implemented by means of collective qubit rotations. In fact, only collective Tavis-Cummings interactions and global qubit rotations are involved. In this way, the total time for the simulation does not scale with the size of the system $N$.
The Dicke model can be investigated provided enough coherence and low-enough gate errors. Notice that this kind of quantum simulation is well suited for superconducting circuits, since simultaneous single-qubit addressing is possible.
Making use of the results in Refs.~\cite{Berry07,Wiebe11}, we demonstrate that the quantum resources needed to approximate the Dicke Hamiltonian with an error less than $\epsilon$ scale efficiently with the number of spins $N$ and of excitations allowed in the bosonic mode $M$. In a Dicke model simulation, one can bound the number of gates $N_\epsilon$ necessary to achieve a certain error $\epsilon$ in a time $t$ by
\begin{equation}
\label{Ngates}
N_\epsilon\leq\frac{2\cdot5^{2k}\left\{2t[\omega_r^RM+N (\omega_q^R+2|g^R|\sqrt{M+1})]\right\}^{1+1/2k}}{\epsilon^{1/2k}}.
\end{equation}  
Here, we have used an upper bound for the norm of the Dicke Hamiltonian, $||H_R||\leq\omega_r^RM+N(\omega_q^R+2|g^R|\sqrt{M+1})$, where $M$ is a truncation on the number of bosonic excitations involved in the dynamics. The fractal depth is set to $k=1$ in the standard Trotter approximations. Using higher orders of fractal decompositions would be a more involved task for implementation of digital approximations in realistic devices, due to the sign inversion that appears~\cite{Suzuki90}. Nevertheless, unitary approximants with arbitrarily high fidelity can be obtained even when $k=1$. 
The formula in Eq.~(\ref{Ngates}) gives an upper bound to the scaling of quantum resources and experimental errors in a simulation involving several qubits. In fact, if one considers a small error for each gate, the accumulated gate error grows linearly with the number of gates.  

Notice that the quantum dynamics of the Dirac Hamiltonian emerges as a specific case of the quantum Rabi dynamics. For the 1+1 dimensional case the algebra of the Dirac spinors $\ket{\Psi}$ corresponds to that of Pauli matrices, and the Dirac equation in the standard representation can be written 
\begin{equation}
\label{Dirac equation 1+1}
i  \frac{d}{d t} \ket{\Psi} = (mc^2 \sigma_z + c p \sigma_x)\ket{\Psi},
\end{equation}
where $m$ is the mass of the particle, $c$ is the speed of light and $ p \propto (a - a^{\dagger}) / i$ is the one-dimensional momentum operator. The Dirac Hamiltonian in Eq.~(\ref{Dirac equation 1+1}), $H_{\rm D}=mc^2 \sigma_z + c p \sigma_x$, shows the same mathematical structure as the quantum Rabi Hamiltonian, Eq.~(\ref{RabiHam}), when $\omega_r^R=0$. This condition can be achieved by choosing $\tilde{\omega}=\omega_r$. The analogy is complete by relating $mc^2$ to $\omega^R_q/2$, $c$ to $g^R$, and the momentum to the quadrature of the microwave field, which can be measured with current microwave technology~\cite{DiCandia}. Choosing an initial state with components in both positive and negative parts of the Dirac spectrum will allow the measurement of the {\it Zitterbewegung}~\cite{Lamata07,Gerritsma10}. By retrieving different quadratures of the microwave field, one can detect this oscillatory motion of the simulated particle in the absence of forces, and the Klein paradox, where a relativistic particle can tunnel through high-energy barriers. To detect such effects, one will be interested in measuring either the position or the momentum of the particle, standing for different quadratures of the microwave field. 

In conclusion, we have shown that the dynamics of the quantum Rabi and Dicke models can be encoded in a circuit QED setup using an analog-digital approach. These quantum simulations will contribute to the observation of quantum dynamics not accessible in current experiments.

\section*{Author Contributions}
A.M. did the calculations and the numerical analysis in Fig. 2. U.L.H. did the numerical analysis in Fig. 3. A.M., U.L.H., J.S.P., L.D., E.S. and L.L. contributed to the developing of the ideas, obtention of the results and writing of the manuscript.

\section*{ACKNOWLEDGMENTS}
 This work is supported by Basque Government IT472-10 Grant, Spanish MINECO FIS2012-36673-C03-02, Ram\'on y Cajal Grant RYC-2012-11391, UPV/EHU UFI 11/55, an UPV/EHU PhD grant, CCQED, PROMISCE and SCALEQIT European projects. UPV/EHU Project No. EHUA14/04.
\section*{ADDITIONAL INFORMATION} 
The authors declare no competing financial interests.

\bibliographystyle{apsrev}

\begin{thebibliography}{99}

 
\bibitem{Rabi36}
Rabi, I. I. 
 \newblock{On the Process of Space Quantization}
\newblock{{\em Phys. Rev.} {\bf 49}, 324 (1936).}


\bibitem{Dicke54} 
Dicke, R. H. 
\newblock{Coherence in Spontaneous Radiation Processes.}
\newblock{{\em Phys. Rev.} {\bf 93}, 99 (1954).}

\bibitem{Jaynes63} Jaynes, E. T.  \& Cummings,  F. W.
\newblock{Comparison of quantum and semiclassical radiation theories with application to the beam maser.}
\newblock{{\em Proc. IEEE} {\bf 51}, 89 (1963).}

\bibitem{Tavis68} Tavis,  M. \& Cummings, F. W.
\newblock{Exact Solution for an $N$-Molecule-Radiation-Field Hamiltonian.}
\newblock{{\em Phys. Rev.} {\bf 170}, 379 (1968).}

\bibitem{Braak11} Braak, D.
\newblock{Integrability of the Rabi Model.}
\newblock{{\em Phys. Rev. Lett.} {\bf 107}, 100401 (2011).}

\bibitem{BraakDicke3}Braak, D.
\newblock{Solution of the Dicke model for N=3.}
\newblock{{\em J. Phys. B} {\bf 46}, 224007 (2013).}

\bibitem{Gunter09} G\"unter, A. et al.
\newblock{Sub-cycle switch-on of ultrastrong light-matter interaction.}
\newblock{{\em Nature} {\bf 458}, 178 (2009).}

\bibitem{Niemczyk10} Niemczyk, T. et al. \newblock{Beyond the Jaynes-Cummings model: circuit QED in the ultrastrong coupling regime.}
\newblock{{\em Nat. Phys.} {\bf 6}, 772 (2010).}

\bibitem{FornDiaz10} Forn-D\'iaz, P. et al.
\newblock{Observation of the Bloch-Siegert Shift in a Qubit-Oscillator System in the Ultrastrong Coupling Regime.}
\newblock{{\em Phys. Rev. Lett.} {\bf 105}, 237001 (2010).}

\bibitem{Casanova10} Casanova, J., Romero, G., Lizuain, I., Garc\'ia-Ripoll, J. J. \& Solano, E.
\newblock{Deep Strong Coupling Regime of the Jaynes-Cummings Model.}
\newblock{ {\em Phys. Rev. Lett.} {\bf 105}, 263603 (2010).}

\bibitem{Crespi12} Crespi, A.,  Longhi, S. \& Osellame, R.
\newblock{Photonic Realization of the Quantum Rabi Model.}
\newblock{{\em Phys. Rev. Lett.} {\bf108}, 163601 (2012).}

\bibitem{Ballester12} Ballester, D., Romero, G., Garc\'ia-Ripoll, J. J., Deppe, F.  \& Solano, E.
\newblock{Quantum Simulation of the Ultrastrong-Coupling Dynamics in Circuit Quantum Electrodynamics.}
\newblock{ {\em Phys. Rev. X} {\bf 2}, 021007 (2012).}

\bibitem{Feynman82}  Feynman., R. P.
\newblock{Simulating Physics with computers.}
\newblock{{\em Int. J. Theor. Phys.} {\bf 21}, 467 (1982).}

\bibitem{Hepp73} Hepp, K. and Lieb, E. H. 
\newblock{On the superradiant phase transition for molecules in a quantized radiation field: the Dicke maser model.}
\newblock{ {\em Ann. Phys. NY }{\bf 76}, 360 (1973).}

\bibitem{Wang73} Wang, Y. K. \& Hioe, F. T. 
\newblock{Phase Transition in the Dicke Model of Superradiance.}
\newblock{{\em Phys. Rev. A} {\bf 7}, 831 (1973).}

\bibitem{Carmichael73} Carmichael, H. J., Gardiner,  C. W.  \& Walls, D. F.
\newblock{Higher order corrections to the Dicke superradiant phase transition.}
\newblock{{\em Phys. Lett. A} {\bf 46}, 47 (1973).}

\bibitem{Bloch12} Bloch, I., Dalibard, J. \& Nascimb\`ene,  S.
\newblock{Quantum simulations with ultracold quantum gases.}
\newblock{{\em  Nat. Phys. }{\bf 8}, 267 (2012).}

\bibitem{Blatt12} Blatt, R. \& Roos, C. F. 
\newblock{Quantum simulations with trapped ions.}
\newblock{{\em Nat. Phys.} {\bf 8}, 277 (2012).}

\bibitem{Alan12} Aspuru-Guzik, A. \& Walther, P.
\newblock{Photonic quantum simulators.}
\newblock{{\em Nat. Phys.} {\bf 8}, 285 (2012).}

\bibitem{Georgescu} Georgescu, I. M., Ashhab, S. \& Nori, F.
\newblock{Quantum simulation.}
\newblock{{\em Rev. Mod. Phys.} {\bf 86}, 153 (2014).} 

\bibitem{Lloyd96}Lloyd, S. 
\newblock{Universal quantum simulators.}
\newblock{{\em  Science} {\bf 273}, 1073 (1996).}

\bibitem{Lanyon11} Lanyon, B. P. et al. 
\newblock{Universal digital quantum simulation with trapped ions.}
\newblock{{\em Science} {\bf 334}, 57 (2011).}

\bibitem{Casanova12}Casanova, J., Mezzacapo, A., Lamata, L. \& Solano, E. 
\newblock{Quantum Simulation of Interacting Fermion Lattice Models in Trapped Ions.}
\newblock{{\em Phys. Rev. Lett.} {\bf 108}, 190502 (2012).}

\bibitem{Mezzacapo12}Mezzacapo, A., Casanova, J., Lamata, L. \& Solano, E.
\newblock{Digital Quantum Simulation of the Holstein Model in Trapped Ions.}
\newblock{{\em Phys. Rev. Lett.} {\bf 109}, 200501 (2012).}

\bibitem{Heras}Las Heras, U. et al.
\newblock{Digital Quantum Simulation of Spin Systems in Superconducting Circuits.}
\newblock{{\em Phys. Rev. Lett.} {\bf 112}, 200501 (2014).}

\bibitem{Koch07}Koch, J. et al. 
\newblock{Charge insensitive qubit design derived from the Cooper pair box.}
\newblock{{\em Phys. Rev. A} {\bf 76}, 042319 (2007).}

\bibitem{Blais04} Blais, A., Huang,  R.-S., Wallraff, A., Girvin, S. M. \& Schoelkopf, R. J.
\newblock{Cavity quantum electrodynamics for superconducting electrical circuits: An architecture for quantum computation.}
\newblock{ {\em Phys. Rev. A} {\bf 69}, 062320 (2004).}

\bibitem{Pedernales13} Pedernales, J. S., Di Candia,  R., Ballester, D. \& Solano, E.
\newblock{Quantum Simulations of Relativistic Quantum Physics in Circuit QED.}
\newblock{ {\em New J. Phys.} {\bf 15}, 055008 (2013). }

\bibitem{Blais07} Blais, A. et al. 
\newblock{Quantum information processing with circuit quantum electrodynamics.}
\newblock{ {\em Phys. Rev. A} {\bf 75}, 032329 (2007).}

\bibitem{Suzuki90}Suzuki, M.
\newblock{Fractal decomposition of exponential operators with applications to many-body theories and Monte Carlo simulations.}
\newblock{ {\em  Phys. Lett. A} {\bf 146}, 319 (1990).}

\bibitem{Berry07} Berry, D. W., Ahokas, G., Cleve, R. \& Sanders, B. C. 
\newblock{Efficient Quantum Algorithms for Simulating Sparse
Hamiltonians.}
\newblock{ {\em Commun. Math. Phys.} {\bf 270}, 359 (2007).}

\bibitem{Wiebe11} Wiebe, N., Berry, D. W., H\o yer, P.  \& Sanders, B. C. 
\newblock{Simulating quantum dynamics on a quantum computer.}
\newblock{ {\em J. Phys. A} {\bf 44}, 445308 (2011).}

\bibitem{DiCandia} Di Candia, R. et al.
\newblock{Dual-path methods for propagating quantum microwaves}
\newblock{ {\em New J. Phys.} {\bf 16}, 015001 (2014).}

\bibitem{Lamata07} Lamata, L., Le\'on, J., Sch\"atz, T.  \& Solano, E. 
\newblock{Dirac Equation and Quantum Relativistic Effects in a Single Trapped Ion.} 
\newblock{{\em Phys. Rev. Lett.} {\bf 98}, 253005 (2007).}

\bibitem{Gerritsma10} Gerritsma, R. et al.
\newblock{Quantum simulation of the Dirac equation.}
\newblock{ {\em Nature} {\bf 463}, 68 (2010).}

\end{thebibliography}

\end{document}